\documentclass[a4paper,twocolumn,11pt,accepted=2022-06-17]{quantumarticle}
\pdfoutput=1
\usepackage[utf8]{inputenc}
\usepackage[english]{babel}
\usepackage[T1]{fontenc}
\usepackage{amsmath}
\usepackage{hyperref}
\usepackage{physics}
\usepackage{tikz}
\usepackage{lipsum}
\usepackage{soul}

\PassOptionsToPackage{compress}{natbib}
\usepackage[numbers]{natbib}
\begin{document}

\title{Deep Reinforcement Learning for Quantum State Preparation with Weak Nonlinear Measurements}

\author{Riccardo Porotti}
\orcid{https://orcid.org/
0000-0002-7632-6141 }
\email{riccardo.porotti@mpl.mpg.de}
\affiliation{Max Planck Institute for the Science of Light, Erlangen, Germany}
\affiliation{Department of Physics, Friedrich-Alexander Universität Erlangen-Nürnberg,
Germany}
\author{Antoine Essig}
\affiliation{Univ Lyon, ENS de Lyon, CNRS, Laboratoire de Physique,F-69342 Lyon, France}
\author{Benjamin Huard}
\orcid{ https://orcid.org/
0000-0002-9848-3658}
\affiliation{Univ Lyon, ENS de Lyon, CNRS, Laboratoire de Physique,F-69342 Lyon, France}
\author{Florian Marquardt}
\orcid{ https://orcid.org/
0000-0003-4566-1753 }
\affiliation{Max Planck Institute for the Science of Light, Erlangen, Germany}
\affiliation{Department of Physics, Friedrich-Alexander Universität Erlangen-Nürnberg,
Germany}

\maketitle

\begin{abstract}
Quantum control has been of increasing interest in recent years, e.g. for tasks like state initialization and stabilization. Feedback-based strategies are particularly powerful, but also hard to find, due to the exponentially increased search space. Deep reinforcement learning holds great promise in this regard. It may provide new answers to difficult questions, such as whether nonlinear measurements can compensate for linear, constrained control. Here we show that reinforcement learning can successfully discover such feedback strategies, without prior knowledge. We illustrate this for state preparation in a cavity subject to quantum-non-demolition detection of photon number, with a simple linear drive as control. Fock states can be produced and stabilized at very high fidelity. It is even possible to reach superposition states, provided the measurement rates for different Fock states can be controlled as well.
\end{abstract}

\section{Introduction}
Modern quantum technologies in their various incarnations, ranging from sensing to computation, rely on quantum control as an essential part of their toolbox. The numerical optimization of pulse sequences, using powerful methods such as GRAPE \cite{khanejaOptimalControlCoupled2005, defouquieresSecondOrderGradient2011}, has by now become an essential aspect of many experiments. Although important challenges remain in this domain, e.g. in multi-qubit control, there is another frontier where the optimization of control strategies is comparatively much less developed: the regime of feedback-based control
\cite{dohertyFeedbackControlQuantum1999,bushevFeedbackCoolingSingle2006,wisemanQuantumMeasurementControl2009,gillettExperimentalFeedbackControl2010,sayrinRealtimeQuantumFeedback2011,campagne-ibarcqPersistentControlSuperconducting2013,ofekExtendingLifetimeQuantum2016,rossiMeasurementbasedQuantumControl2018,hacohen-gourgyContinuousMeasurementsControl2020,fallaniLearningFeedbackControl2022}. The automatic discovery of feedback strategies, which include real-time decision-making based on previously observed measurement results, is challenging for two reasons: First, standard optimal control techniques are not applicable, and second, feedback leads to a drastic expansion of the search space, involving a doubly-exponential growth of the number of strategies vs. the number of time steps.
Deep reinforcement learning (RL) \cite{suttonReinforcementLearningSecond2018} is a general approach that offers a possible solution for this challenge, with powerful recent examples in other areas, ranging from computer games \cite{mnihHumanlevelControlDeep2015} to robotics \cite{haarnojaLearningWalkDeep2019}. Deep RL for discovering quantum feedback strategies was first introduced in \cite{foselReinforcementLearningNeural2018} where it was illustrated in the application of quantum error correction. The idea behind deep RL is that a neural network-based "agent" makes observations and suggests corresponding actions. During the learning phase, the agent discovers from scratch novel strategies, in a systematic procedure which at first resembles trial and error but later begins to build on insights acquired earlier. That RL in general and deep RL in particular is a powerful approach in quantum physics has by now been demonstrated in a variety of tasks in different areas: in most works so far, these tasks did not yet require real-time feedback involving decision-making based on physical measurements, but RL already proved itself to be a versatile tool even in those settings \cite{chenFidelityBasedProbabilisticQLearning2014,augustTakingGradientsExperiments2018a,bukovReinforcementLearningDifferent2018,porottiCoherentTransportQuantum2019,niuUniversalQuantumControl2019,anDeepReinforcementLearning2019,xuGeneralizableControlQuantum2019,arrazolaMachineLearningMethod2019,odriscollHybridMachineLearning2019,foselEfficientCavityControl2020,dalgaardGlobalOptimizationQuantum2020b, maCurriculumbasedDeepReinforcement2021,anQuantumOptimalControl2021, baumExperimentalDeepReinforcement2021, foselQuantumCircuitOptimization2021}. These publications are part of a larger drive towards the use of machine learning tools for quantum experiments (e.g. \cite{flurinUsingRecurrentNeural2020,lennonEfficientlyMeasuringQuantum2019, jungDeepLearningEnhanced2021, nguyenDeepReinforcementLearning2021}). A small number of works have explored the use of various RL(-related) techniques to optimize feedback or adaptive measurement protocols \cite{hentschelMachineLearningPrecise2010,tierschAdaptiveQuantumComputation2015,palittapongarnpimLearningQuantumControl2017,mackeprangReinforcementLearningApproach2020}. Investigations of deep RL, which promises to be a powerful approach in the domain of measurement-based quantum feedback, are still rare. Beyond \cite{foselReinforcementLearningNeural2018}, there have been some very interesting additional works in this direction. These include optomechanical feedback cooling \cite{sommerProspectsReinforcementLearning2020} as well as state preparation and stabilization in a potential with position measurements \cite{wangDeepReinforcementLearning2019, borahMeasurementBasedFeedback2021}.  Very recently, feedback-based quantum state preparation in a cavity-qubit system \cite{sivakModelFreeQuantumControl2021} was proposed, for a setting employing continuously parameterized quantum circuits.
\begin{figure}[t]
	\includegraphics[width=\linewidth]{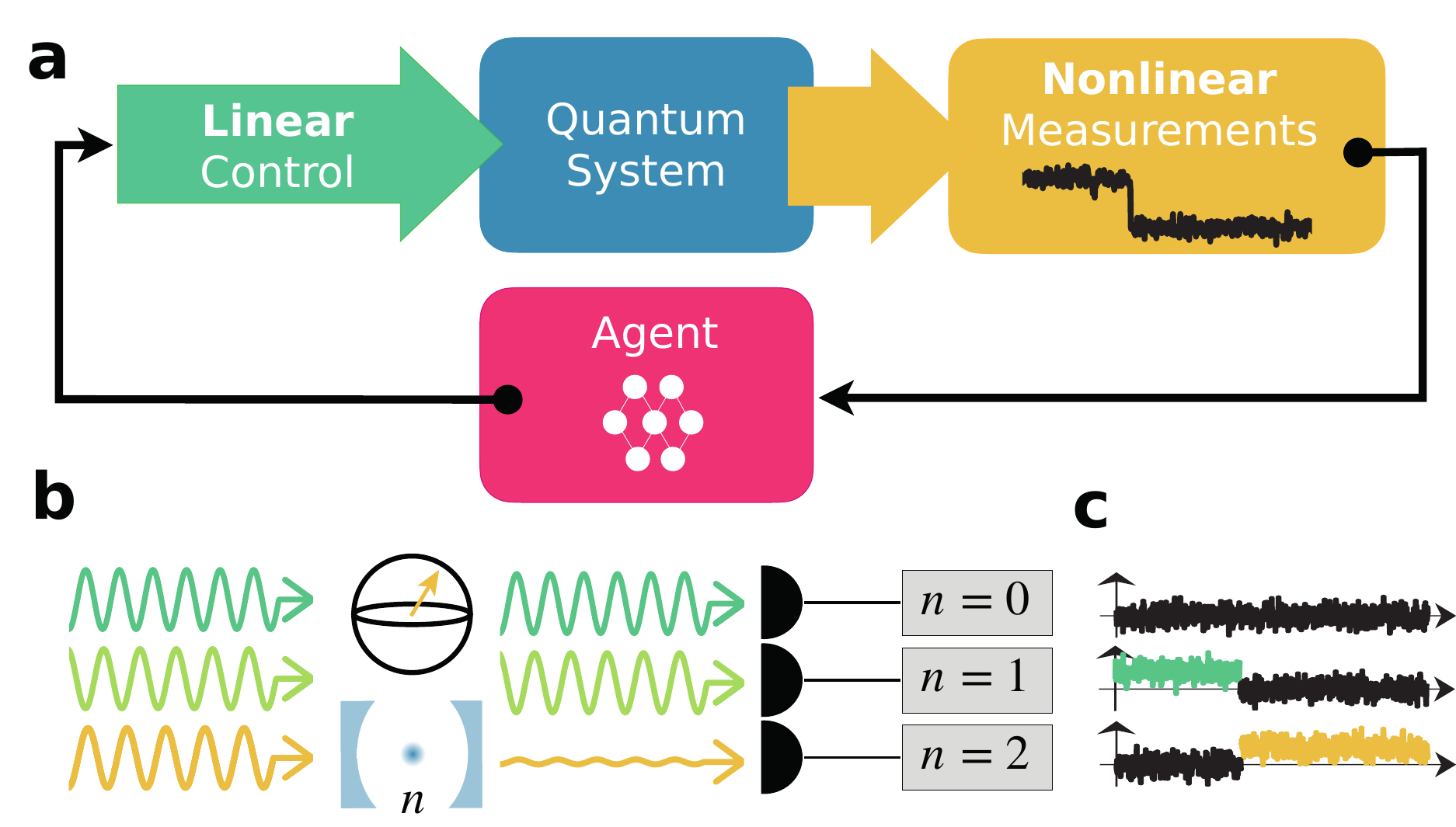} 
	\caption{Reinforcement learning for discovering feedback control based on continuous measurements. a) A neural-network-based agent can find a strategy to control a quantum system, compensating for simple (linear) controls by extracting information from advanced (nonlinear) measurements. b) Schematics of the model considered in the main text (see \cite{essigMultiplexedPhotonNumber2020}): a qubit is illuminated with multiple frequencies in order to probe the number of photons $n$ in a coupled cavity. The information can be extracted by observing the reflected radiation in each of the channels. The reduction in amplitude in the outgoing wave indicates detection of the Fock state corresponding to the respective channel. c) Timetraces of the quadrature signal observed in the homodyne measurement channels of Eq.~\ref{eq:homodyne}. In this particular example, the cavity state decays at an intermediate time.
	}
	\label{fig:figure1}
\end{figure}
Deep RL for measurement-based quantum feedback can be used to address experimentally relevant questions where no general straightforward answer is available. While general results exist in the absence of feedback, e.g. which terms in a Hamiltonian need to be controllable to produce arbitrary unitaries, such insights are not available for feedback-based strategies. One obviously important example concerns the question whether one can compensate for simple control (e.g. purely linear operations) by exploiting more advanced nonlinear measurements (Fig.~\ref{fig:figure1}a). In the present work, we answer this question by investigating an example inspired by recent experimental progress: a harmonic oscillator where individual Fock states can be continuously monitored in a quantum non-demolition fashion. The corresponding experimental setting (demonstrated in \cite{essigMultiplexedPhotonNumber2020}) concerns a microwave cavity where the precise Fock state number can be monitored by driving a coupled qubit with a frequency comb. 

Using a state-of-the-art deep RL approach, we show that under these conditions a mere displacement drive, conditioned on the measurement outcome, is sufficient to produce and stabilize Fock states. Even superpositions of such states can be stabilized to a good degree of fidelity, provided the control is expanded to include adaptive measurement rates.

\section{Physical system}

Measurement-based feedback for quantum physics, using digital controllers, dates back to 2011 with a pioneering experiment that demonstrated the stabilization of a Fock state in a microwave cavity using the information extracted by interacting flying circular Rydberg atoms \cite{sayrinRealtimeQuantumFeedback2011}. Each circular Rydberg atom encodes one bit of information about the number of photons in the cavity. From the extracted information, a classical controller computes the density matrix of the cavity in real time. It can then react in one of two ways. Either it sends a resonant atom that can add or subtract a photon to the cavity \cite{peaudecerfQuantumFeedbackExperiments2013,zhouFieldLockedFock2012}, or it sends a classical microwave drive whose amplitude is chosen to displace the cavity state in an optimal way \cite{sayrinRealtimeQuantumFeedback2011}. The latter case makes a fascinating use of quantum measurement backaction as it is the combined effect of a classical drive and quantum measurement that steers the cavity towards the targeted Fock state. We will propose a more general scheme, that can accommodate the preparation of superpositions, or that can deal with the extra presence of noise, without the need of constructing the strategy by hand. 

Several experiments with Rydberg atoms or superconducting circuits have by now been realized to count the number of photons in a microwave resonator \cite{curtisSingleshotNumberresolvedDetection2021,guerlinProgressiveFieldstateCollapse2007,johnsonQuantumNondemolitionDetection2010,peaudecerfAdaptiveQuantumNondemolition2014}, enabling them in principle to implement quantum feedback strategies of this sort. 

The present work is heavily inspired by a recent implementation~\cite{essigMultiplexedPhotonNumber2020} which demonstrated a way to acquire information about the number of photons in a resonator in a continuous manner. It exploits a qubit, coupled to both the cavity and a measurement transmission line, driven by multiple frequencies simultaneously, each of which is sensitive (via dispersive qubit-cavity coupling) to one Fock state. 

Thus, in this experiment, to each number of photons $n$ in the cavity is associated one of the frequencies of the frequency comb. The signal at this particular frequency is affected (i.e. the coherent state is displaced) when exactly $n$ photons are present in the cavity. This situation is shown in Fig.~\eqref{fig:figure1}b. 

If we change the reference frame by displacing the state by the opposite of the input coherent state (such that the input state would turn into the ground state), we can equivalently say the following about the physical situation in the experiment: for any given cavity photon number $n$, only the field at a single frequency is excited and all the other frequencies are in the vacuum state. 

This measurement introduces an unavoidable intrinsic fluctuating back-action but also tries to localize the system into some random Fock state according to the Born rule - not necessarily the desired target state. Any kind of control must both counteract and exploit these tendencies introduced by the measurement \cite{peaudecerfQuantumFeedbackExperiments2013}.

\section{Model}

We now design a model for this situation that is as simple as possible, without going into the dynamics of the propagating waves. The basic idea is that, in our model, we replace the travelling waves at the different frequencies of the measurement frequency comb with artificial localized modes, which are driven and can be treated in the usual input-output formalism. Each of these modes than can experience a frequency shift (leading to a phase shift of a wave transmitted through this mode), provided the cavity photon number $n$ takes on the value matching the index of this mode.  

Apart from drive and decay (which we will introduce later), the Hamiltonian of the cavity subject to the Fock state QND measurement can be expressed as
\begin{equation}
\label{eq:H_meas}
{\hat H_{\rm meas}}=\sum_n \chi_n {\hat a_n}^{\dagger}{\hat a_n} {\hat P}_n,
\end{equation}
where ${\hat P}_n=\ket{n}\bra{n}$ is the projector on Fock state $n$ of the cavity and we have switched to a frame rotating at the cavity resonance frequency. The mode ${\hat a}_n$ refers to the $n$-th measurement channel, which will be monitored to assess the cavity state, and which, as we have said above, will be treated like a driven, decaying bosonic mode. We will choose $\chi_n$ equal for all $n$ from now on. Eq.~\eqref{eq:H_meas} represents the simplest possible model that captures the essence of such a multichannel dispersive Fock state measurement.

At this stage of the effective description, we have already eliminated the qubit which is coupled to the cavity and which provides the nonlinearity needed for the dispersive coupling displayed above. 
When a homodyne measurement is performed on the phase quadrature of any of the measurement channels (i.e. the output of any of these localized modes ${\hat a}_n$), the resulting evolution of the cavity's reduced density matrix, obtained after integrating out the measurement channels ${\hat a}_n$, can be described by a stochastic master equation (SME) \cite{gardinerQuantumNoiseHandbook2004,wisemanQuantumMeasurementControl2009}. With ${\hat A}={\hat P}_n$ as the measurement operator, the contribution to the SME from this channel reads:
\begin{equation}
    \label{eq:SME}
    d {\hat \rho}=\gamma_{\rm meas}\mathcal{D}[\hat A] {\hat \rho}(t) + \sqrt{\gamma_{\rm meas}}\mathcal{H}[\hat A] {\hat \rho}(t)  dW(t)
\end{equation}
where $\gamma_{\rm meas}$ is the measurement rate of a perfectly efficient measurement, $\mathcal{D}[\hat A]$ is the measurement-induced Lindblad superoperator, 
\begin{equation}
   \mathcal{D}[\hat A]{\hat \rho}=\hat A {\hat \rho} \hat A^{\dagger}-\frac{1}{2}\left({\hat \rho} \hat A^{\dagger} \hat A  + \hat A^{\dagger} \hat A {\hat \rho} \right),
\end{equation}
and $\mathcal{H}[\hat A]$ is 
\begin{equation}
\mathcal{H}[\hat A] {\hat \rho}(t)=(\hat A \hat \rho + \hat \rho \hat A - \left<\hat A + \hat A^{\dagger} \right> \hat \rho )
\end{equation}
with $ \left< \hat A  \right>=\mathrm{Tr}[\hat A \hat \rho]$.

The stochastic part of Eq.~\eqref{eq:SME} contains $dW(t)$, which is an infinitesimal Wiener process, with zero mean and  $\langle dW(t)^2 \rangle=dt$. The Wiener process is experimentally obtained from the homodyne measurement record, which reads
\begin{equation}
\label{eq:homodyne}
    \sqrt{\frac{\gamma_\mathrm{meas}}{2}}\mathrm{Tr}(\hat{A}\rho+\rho\hat{A}^\dagger)dt+dW(t).
\end{equation}

Combining the effect of all the simultaneous measurements, plus an additional coherent drive, we obtain the Itô stochastic differential equation of motion for the quantum state of the cavity:
\begin{equation}
\label{eq:SME_final}
\begin{split}
d {\hat \rho} = &-i[{\hat H}(t),{\hat \rho}(t)] dt+ \\
&+\sum_n \frac{\gamma_n}{2} \mathcal{D}[{\hat P}_n] \hat \rho dt+\\
&+ \frac{\sqrt{\gamma_n}}{2} ({\hat P}_n \hat \rho + \hat \rho {\hat P}_n - \left< 2 {\hat P}_n  \right> \hat \rho ) dW_n(t)
\end{split}
\end{equation}
where, if not otherwise mentioned, we choose the same rate $\gamma_n=\gamma_{\rm meas} $ for all channels and the Wiener processes $dW_n(t)$ associated to every channel are uncorrelated $\langle dW_n(t)dW_m(t')\rangle=\delta(t-t')\delta_{nm}\mathrm{d}t$. 
Here ${\hat H}(t)$ describes a time-dependent drive that is used to control the cavity conditioned on the measurement outcomes, implementing feedback control. That is, 
\begin{equation}
    H(t)= i \left(\beta(t) {\hat a}^{\dagger}-\beta^{*}(t) {\hat a} \right)
\end{equation}
where $\beta(t)$ is the complex time-dependent drive amplitude.

\begin{figure}[t]
	\includegraphics[width=\linewidth]{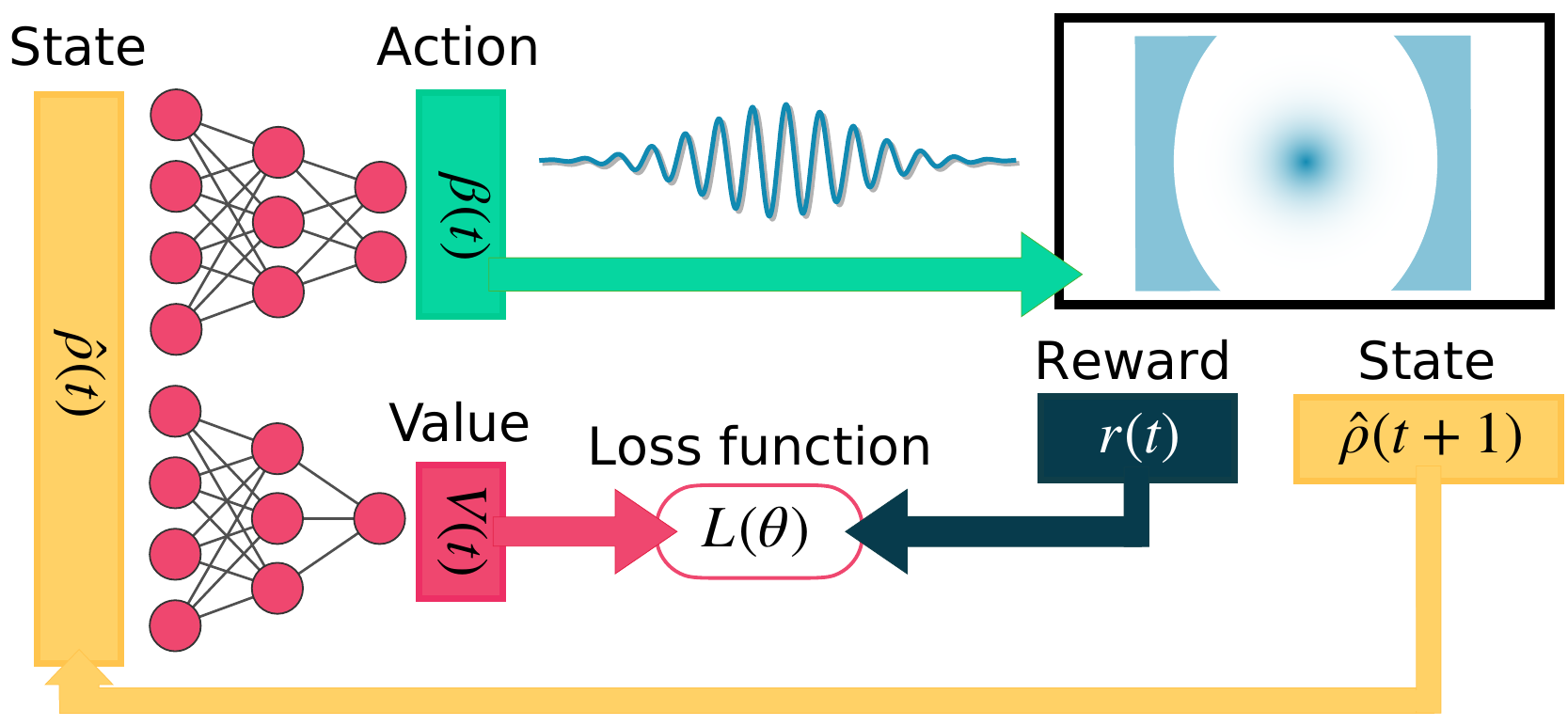} 
	\caption{Reinforcement Learning pipeline.  A fully connected neural network receives the quantum state of the cavity (represented via the density matrix) at timestep $t$. The network outputs the suggested action (i.e. real and imaginary part of the  displacement drive) to be applied to the cavity in that particular timestep. The driven dissipative quantum system is evolved following Eq.~\eqref{eq:SME} in order to obtain the density matrix at the next timestep $t+1$. Furthermore, the instantaneous reward value $r(t)$ is computed from the current state fidelity and enters the loss function $L$. We are using an actor-critic method, in which the current quantum state is also assigned a value function $V(t)$, by another neural network, which helps in searching for the optimum feedback strategy.  }
	\label{fig:figure2}
\end{figure}
\begin{figure}[t!]
\centering
	\includegraphics[width=1.0\linewidth]{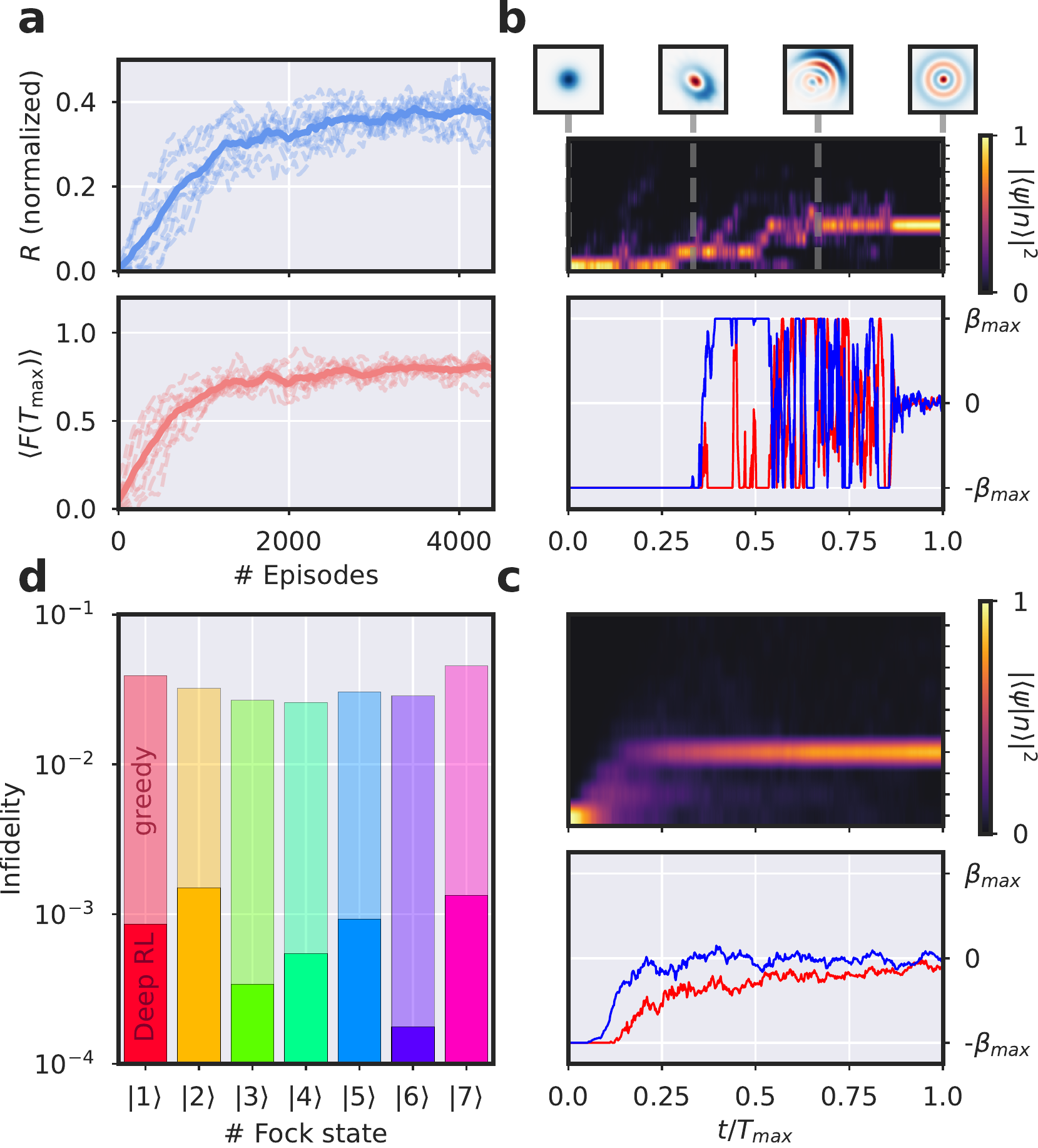} 
	\caption{Reinforcement learning for cavity state preparation: Fock state results. a) Training progress of the RL agent (here, for reaching Fock state $\ket{3}$). Upper panel: Cumulative reward (i.e. the sum of all instantaneous rewards during a trajectory) during different training episodes (the solid line is the average). The return is normalized in the sense that having the maximum reward in each timestep would result in $R=1$. Lower panel: Final state fidelity in each episode (maximum 1.0). b) Example of a trajectory after the training is completed. Upper panel: Fock state probability distribution of the cavity, where the stochasticity stems from the weak Fock state measurement. Wigner densities corresponding to times $[0, 1/3, 2/3, 1]T_{max}$ are shown. Lower panel: Real (blue) and imaginary (red) part of the displacement drive. The actions are capped in range to $\pm\beta_{max}$. c) Same information, but averaged over 50 trajectories.  d) Results after training, in terms of final infidelity $1-F$, averaged over multiple trajectories, displayed for different networks trained on reaching different final target Fock states. Lower infidelities (compared to the training shown in panel a) are reached by turning the policy into a deterministic one, always taking the action corresponding to the peak of the Gaussian distribution suggested by the neural network. Note logarithmic axis. Saturated color represents the RL approach, pale color the greedy strategy (performing much worse). In all examples, $\gamma_{meas}T_{max}=400$ and $\beta_{max}=20$.}
	\label{fig:figure3}
\end{figure}

The main point that we want to stress is that the control is completely linear, while the nonlinear part of Eq.~\eqref{eq:SME} is given by the homodyne measurement only.

The goal that we want to achieve with the help of a suitable feedback-based control strategy is quantum state preparation, i.e. to drive an arbitrary initial state to a prescribed target state.

\section{Reinforcement Learning}

In this section we will show how Reinforcement Learning (RL) can be applied to the problem at hand. In general, Reinforcement Learning tries to discover good feedback-based control strategies. An "agent" observes the world (typically called "environment") and takes an action based on that observation ("state"). The mapping from observation to action defines the strategy ("policy"). Modern powerful RL variants, such as the one we will be using here, produce this mapping with the help of a deep neural network. The aim of RL training is to maximize a reward that has been designed to express the goal of the particular task at hand. This is done via running many thousand trajectories, improving the policy gradually, e.g. via some version of gradient descent on parameters that determine the policy (i.e. typically the parameters defining the agent's network). 
In our case, we want to control the cavity (the RL "environment") via a linear drive, based on the noisy measurement trajectories of the nonlinear measurement. The goal is to prepare some desired quantum state and stabilize it against noise, e.g. noise coming from the measurement itself or from external decay or decoherence.
The RL environment is modeled using a simulation of the physical time evolution of the cavity, employing the SME introduced above, which we solve with a Runge-Kutta method with additive noise. The physical time $t$ is discretized in $N_{\rm max}$ timesteps, such that $t$ can only assume discretized values $t_j=\Delta t \cdot j$ with $j \in[0,N_{\rm max}]$. At each time $t_j$, the environment receives an action $a_j$, carries out this action, and outputs the next observation $s_{j+1}$ and a reward $r_j$. We will define all of these key variables in the following paragraph.

The obvious choice for the observation $s_j$ would be to feed into the RL agent some representation of a noisy measurement trajectory. Then, the network would need to learn to filter the trajectory in a suitable way to learn as much as possible about what is going on inside the quantum device. However, it turns out that (i) solving this challenge simultaneously with the RL problem is surprisingly hard and (ii) we can apply another approach that is more efficient and also works well in an experiment. In fact, to solve this subtask, we can take the noisy measurement traces and use them to evolve the quantum state according to the stochastic ME. This can be performed online in a real experiment (as long as the quantum device is well characterized beforehand, and we have a sufficiently fast signal processing device, e.g. a field-programmable gate array, FPGA). Therefore, we choose to feed the result of this procedure into the RL agent, meaning we provide the quantum state ${\hat \rho}$ as input. Any measurement inefficiencies will be encoded in this quantum state, increasing the uncertainty and decreasing the purity. This approach has been applied successfully before, e.g. in \cite{ porottiCoherentTransportQuantum2019, foselReinforcementLearningNeural2018}.
Alternatively, one can apply the idea of 'two-stage learning' introduced in our previous work \cite{foselReinforcementLearningNeural2018}. In this approach, one uses a supervised learning approach to train a recurrent network, which can be deployed in an experiment as it receives only the measurement results as input. Supervised training is performed by teaching this recurrent network to mimic the output (i.e. the action sequences) produced by the original network, which was fully trained via the RL technique discussed here.

In order to avoid confusion, we note that for our case, there is no distinction between "observation" and "state". Usually an observation can be a partial representation of the state of the system, in which case we have a partially observable Markov decision process (POMDP), or it can contain the full information about the system, and in that case we have a standard Markov decision process (MDP). We will use the density matrix ${\hat \rho}$ as observation, such that the two definitions of state and observation coincide in this case. On the technical level, the state $s$ fed into the neural network's input layer is a vector of length $2N^2$ representing all the real and imaginary parts of the entries of the cavity's density matrix. Here $N$ is the cutoff in the Fock state basis of the cavity that we use in our simulations.

The actions $a_j$ are the real and imaginary part of the displacement drive applied to the cavity: $a_j = [\Re(\beta(t_j) ), \Im(\beta(t_j))]$.

\begin{figure}[t!!!]
\centering
	\includegraphics[width=1.0\linewidth]{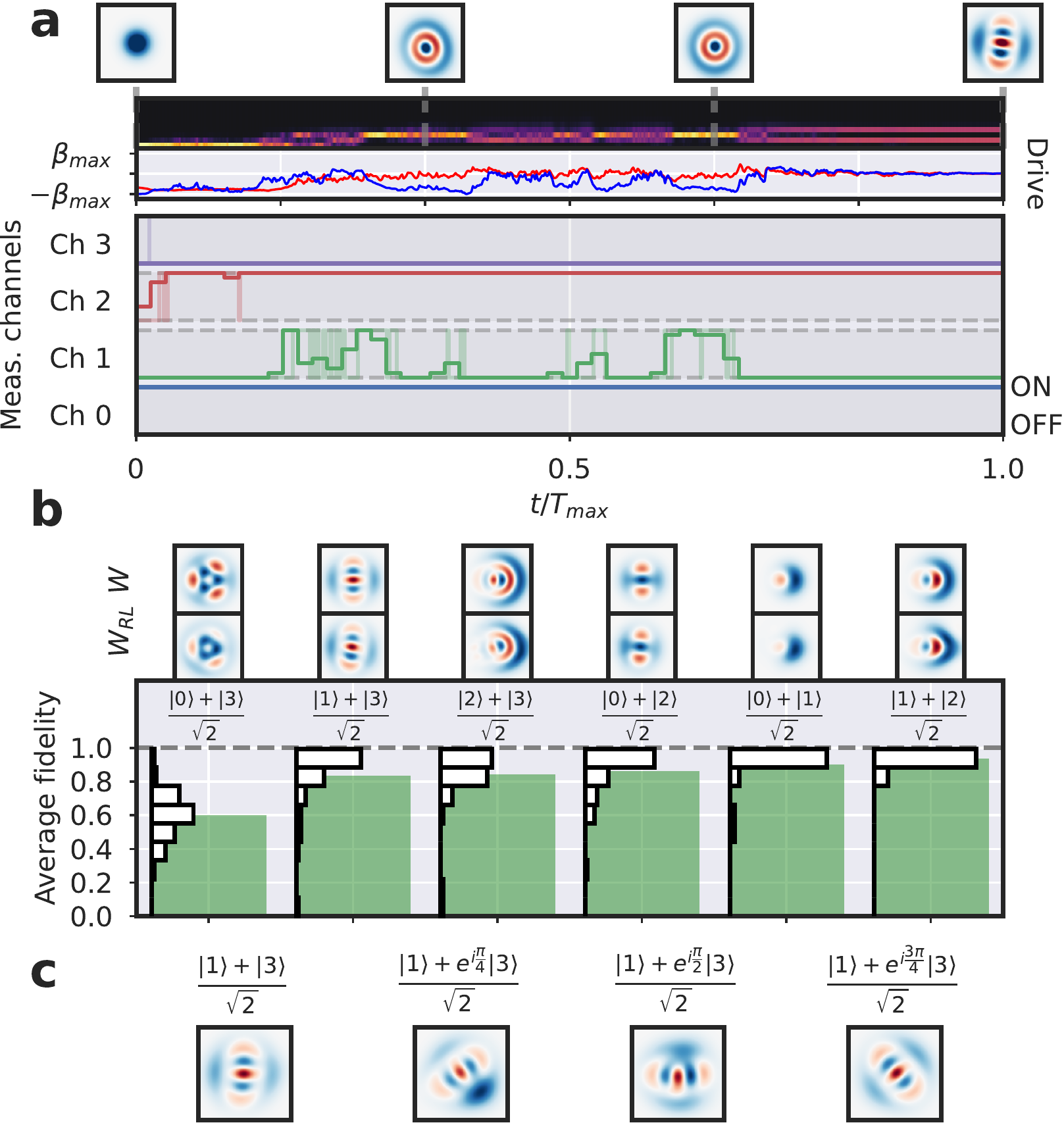} 
	\caption{Preparation of Fock state superpositions, via additional measurement channel control. a) Example of a trajectory for the $\frac{\ket{1}+\ket{3}}{\sqrt{2}}$ target state (probability distribution, top, and drive, bottom). In the four lower plots, the shaded traces represent the ON-OFF control of the measurement channels corresponding to Fock states $\ket{0}$ to $\ket{3}$. The solid lines are a moving average over 10 timesteps. b) Average fidelities reached after the training for different superposition states. For each case, we show the distribution of final fidelities (white histogram, obtained from 50 trajectories) and the resulting average fidelity. Above that panel, we show the comparison between the Wigner densities for the target state (top) and the average state obtained from the strategy (bottom). Smaller distance between the two states forming the superposition can yield better results. c) Average Wigner densities reached by RL, with different relative phases between state $\ket{1}$ and $\ket{3}$. In all cases, we only control the first $M$ measurement channels, with $M=4$. }
	\label{fig:figure4}
\end{figure}
Later, the actions will be extended to accommodate control of the measurement rate in the various channels. The actions are predicted at each timestep by a policy neural network, which receives the current density matrix of the system as its input.

In the version of continuous-action RL we are using here, each neuron in the output layer represents the mean of a Gaussian distribution from which the actions are extracted, while the variance stays fixed. After the predicted value is extracted, the actions get manipulated by the agent according to the particular problem at hand. In our case, the components of the action represent the real and imaginary parts of the displacement drive, but they are limited to a range $[-1,+1]$. We rescale them by a fixed factor $\beta_{\rm mult}$ and the resulting drive is applied to the system for the current time step.

The reward function is a critical choice in RL scenarios. It encourages the RL agent to find a robust policy that, in our case, drives the physical system to the desired target state. In this work, we chose as instantaneous reward a function of the fidelity $F(t)$, computed between the current density matrix ${\hat \rho}(t)$ and the target state ${\hat \rho}_{target}(t)$:
\begin{equation}
r_t = |F(t)|^{\theta}
\end{equation}where the usual definition of quantum fidelity $F(t)=\left( \text{tr}\left [\sqrt{\sqrt{\rho(t)} \rho_{target}\sqrt{\rho(t)}}\right]\right)^2$ is used. We introduced an exponent $\theta > 1$ because we want to punish states which only have a moderately high fidelity (e. g. $F(t)=1/2$ when $\ket{\psi}=\ket{n}$ and $\ket{\psi_{target}}=\frac{1}{\sqrt{2}}(\ket{n}+\ket{m})$ ) in favour of states with fidelity much closer to 1. 
In our numerical experiments, it turned out that this modified reward was crucial to the success of the algorithm. Unless otherwise specified, in the following we will use $\theta = 8$. As usual, the total return $R$ of a trajectory is then the sum of all instantaneous rewards, i.e. $R=\sum_t r_t$.

As mentioned before, the RL agent is modelled with a neural network, which is trained in order to maximize the cumulative reward in a trajectory. We will refer to the network as Policy Network. In actor-critic RL methods, such as the one we will be using here, this network is supplemented by a Value Network, which is used to predict the expected cumulative reward if one starts in the current state, serving as a baseline. Specifically, as our RL approach we decided to use a Proximal Policy Optimization (PPO) algorithm \cite{schulmanProximalPolicyOptimization2017}, which is known to be a modern,  general, easy-to-implement and sample-efficient variant of policy gradient techniques. This algorithm is closely related to Trust Region Policy Optimization (TRPO) \cite{schulmanTrustRegionPolicy2017} techniques, in the sense that they rely on updating the current policy according to some constraints, limiting sudden jumps in the latter (see Fig.\ref{fig:figure2} and the appendix for the general layout of the algorithm). Even though this is a time-dependent control problem, using recurrent neural networks, i.e. networks with memory, is not necessary in our approach: knowing the quantum state at any time gives the maximum amount of information that is available and that can possibly influence the choice of the next action. The output of the Policy Network, for each action, is a Gaussian distribution from which the action is extracted. The deep RL agent, during the training, learns the peak location of the Gaussian while the standard deviation is kept fixed. This strategy is useful during the training to explore the action space, and naturally deals with stochastic environments, like the one in study. RL techniques for efficient control with continuous action spaces are a relatively recent development. For the present work, we perform the training of the RL agent using the Python library Stable Baselines \cite{hillStableBaselines2018}. 

\begin{figure}[t]
\centering
	\includegraphics[width=1.0\linewidth]{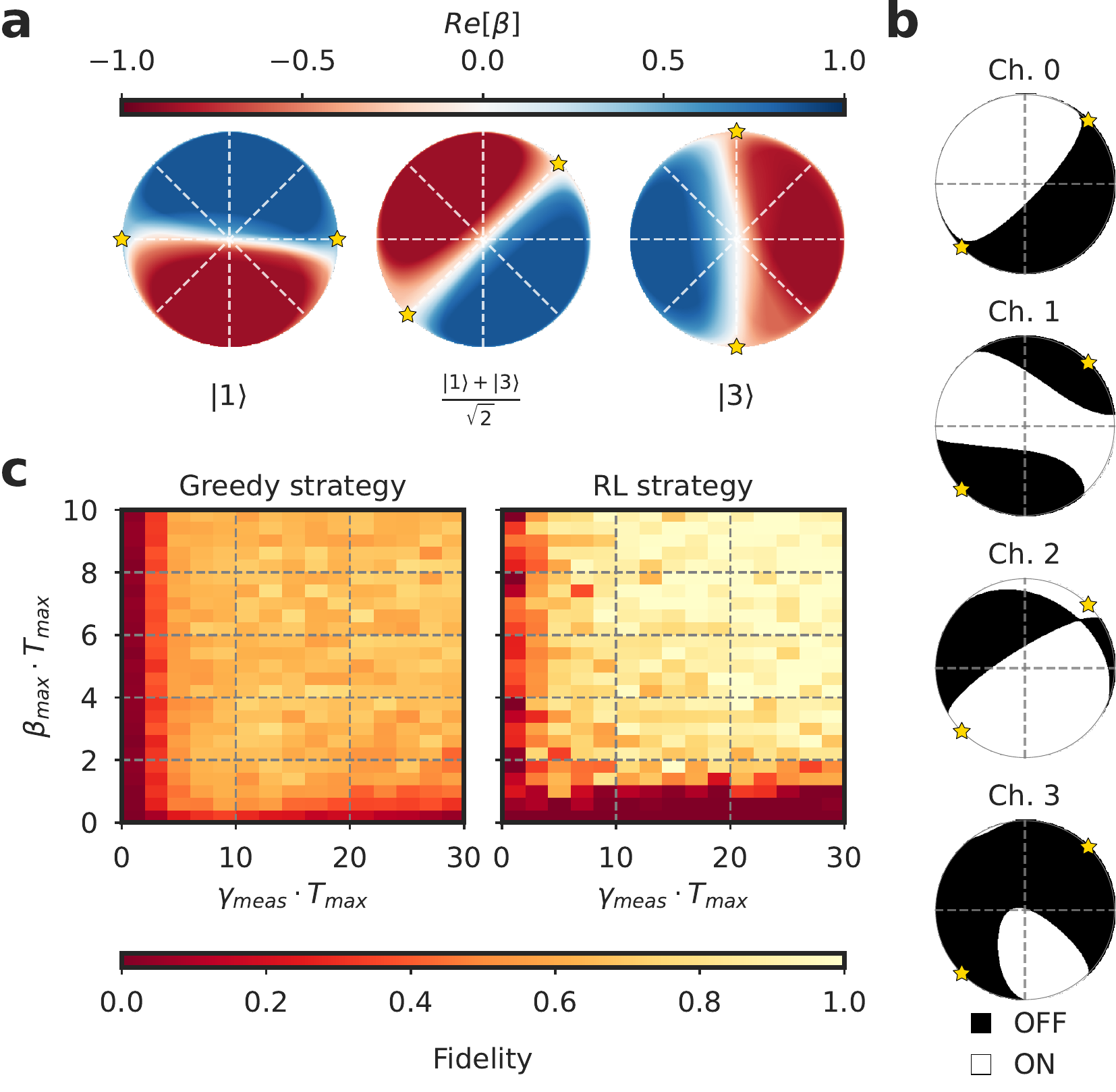} 
	\caption{
	Analysis of the strategy discovered by deep RL. a) Response of different neural networks trained on target states $\ket{1}$ ,$\frac{\ket{1}+\ket{3}}{\sqrt{2}}$, $\ket{3}$, respectively. To visualize the output of the neural network as a function of its input, the input state is parameterized  with two variables $(x,y)$:  $x\ket{1} + \sqrt{1-x^2-y^2 } \ket{2} + y\ket{3}$; thus the stars in each plot represent the corresponding target state (up to a global phase). For each panel, we show the real value of the displacement drive $\beta(t)$, while the imaginary part is negligible (averaging the outputs of 7 different converged networks). b) The measurement channel actions discovered for the $\frac{\ket{1}+\ket{3}}{\sqrt{2}}$ target state, plotted in the same way as before vs parametrized input state.  c) Average final fidelity as a function of measurement rate $\gamma_{\rm meas}$ and maximum drive amplitude $\beta_{\rm max}$ (for the example $\ket{\psi_{\rm target}}=\ket{3}$). Comparison between the greedy strategy (see main text) and Reinforcement Learning. RL is more robust in a wider range of physical parameters of the system.
    }
	\label{fig:figure5}
\end{figure}
\section{Results}

Many different stochastic trajectories (or "episodes"), all starting from the ground state $\ket{\psi}=\ket{0}$, are run in parallel, then the data are collected, and the two networks, the policy and the value neural network, are updated. Unless otherwise specified, in our numerical experiments each trajectory is made up of 1000 timesteps. At each timestep, the agent inputs the current density matrix to the policy neural network, it extracts real and imaginary parts of the displacement, and it applies that displacement drive to the cavity.

In Fig.~\ref{fig:figure3}, we show the main results for our RL agent that was trained on reaching different Fock states, from $\ket{\psi_{target}}=\ket{1}$ to $\ket{\psi_{target}}=\ket{7}$. All of these results are still in the absence of decay or dephasing, except for the unavoidable, intrinsic measurement-induced dephasing processes described by the SME. We will later return to verify the performance of the RL approach when this restriction is relaxed.

Starting from the ground state, the network learns to displace to reach higher energies. At the same time, the measurement has two tendencies: it introduces fluctuations but it also tries to collapse onto some random Fock state (not necessarily the desired target state!). The network eventually finds a strategy to compete with and exploit the measurement process, ending up in the target state and stabilizing that state. The RL agent reaches high fidelities successfully after seeing only about 2000 trajectories, which is surprisingly efficient in the context of deep RL. In our case we explored quantum state preparation until a target Fock state number $7$ without any noticeable loss of fidelity.

A typical feedback control scheme discovered by the network is shown in Fig.~\ref{fig:figure3}b,c. The goal is to reach a Fock state from the ground state. At first the network applies a deterministic displacement drive of maximal amplitude both in the real and imaginary parts of the drive. This is because that allows it to reach higher energies of the cavity most quickly. After a while, when higher excitations have been reached, the stochastic outcomes of measurements force the agent to respond in a way that depends on those outcomes (giving rise to a stochastically fluctuating control trajectory). Finally, after the correct Fock state has been reached, the control amplitude is reduced to zero.

The resulting strategies obtained by the agent surpass even the best benchmark strategy that we could construct based on our physical intuition, a greedy strategy sketched in Appendix A. Essentially, this greedy strategy attempts, at each time, to choose a displacement that would maximize the overlap with the target state.
\\

Now we move on to a more challenging problem, where the target states become superpositions of Fock states. This is in principle incompatible with continuously active Fock state measurements, because these will ultimately destroy any such superposition. We therefore allow an increase in capabilities for the nonlinear measurement, enabling the RL agent to control the measurement rates for the various channels. This is possible in a real experiment by controlling the amplitudes of the microwave tones applied to the system. In principle, the RL algorithm could handle without any problems the extension to $M$ additional continuous actions, representing the measurement rates. We decide to go for a simplified approach, in order to better understand the resulting strategy after the training: in this simpler approach, $\chi_n$ of Eq.~\eqref{eq:H_meas} can only be 0 or 1, meaning that each of the first $M$ channels (corresponding to state $\ket{0}$ to $\ket{M-1}$ are respectively either OFF or ON at any moment of time. Therefore, in Fig.~\ref{fig:figure4} the actions are $M+2$ dimensional: real and imaginary part of the displacement drive and $M$ ON-OFF measurement rates. In Fig.~\ref{fig:figure4}a we visualize the strategy chosen by the RL agent, for the target state $(\ket{1}+\ket{3})/\sqrt{2}$. In the beginning of the trajectory, the values of the measurement rates are not critical, so the agent just learns to displace the ground state with the maximum displacement available. Then, in the middle of the trajectories, the strategy relies on a counter-play between the measurement-induced collapse of the state and the effect of the drive. The agent finds out that the best strategy is to keep channels 0 and 2 ON (measurements on Fock states $\ket{0}$ and $\ket{2}$) while keeping channel 3 completely OFF. Instead, channel 1 is kept OFF when the state has collapsed to the correct one, while on average is ON in the middle of the trajectory, possibly in order to avoid the broadening of the state. In Fig.~\ref{fig:figure4}b, the final states reached are plotted in comparison to the true target states. For superpositions between Fock states that are further apart, it is harder to reach large fidelities. However, it is worthwhile to notice that only the first 4 channels are controlled in these examples, in addition to the already limited linear drive controls. From this perspective, it is still remarkable how well even superpositions can be produced (as also evidenced by the Wigner densities).

\section{Analysis and robustness of the network }

\begin{figure}[t]
\centering
	\includegraphics[width=1.0\linewidth]{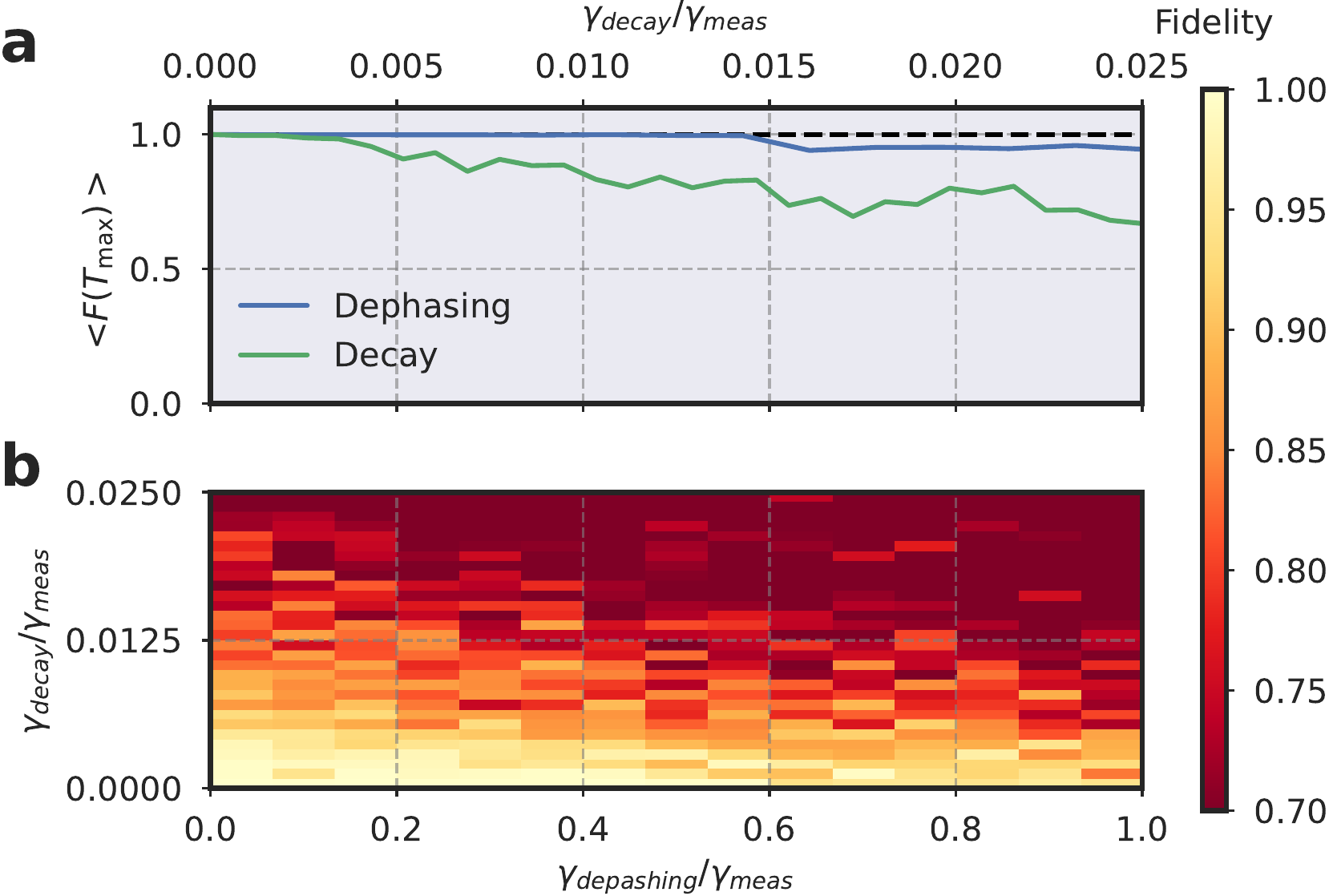} 
	\caption{
Robustness of the Reinforcement Learning approach vs. decay and dephasing (or measurement inefficiency). The target state in this example is $\ket{3}$. a) Average final fidelity vs either dephasing rate or decay rate; the dashed line represents the fidelity without noise (compare Fig.~\ref{fig:figure3}). Each data point represents the strategy of a freshly trained neural network. b) Average fidelity when both decay and dephasing are present, in a 2D parameter sweep.
    }
	\label{fig:figure6}
\end{figure}
In order to analyze the behaviour of the trained agent, we study the relation between input (state) and output (action) that the policy network has learned after the training is completed. Our first analysis inquires how the network trained on a particular target state behaves when it receives a state which slightly differs from the target. In Fig.~\ref{fig:figure5}a we show the input-output relations of three different trained neural networks, trained to reach states $\ket{1}$ ,$\frac{\ket{1}+\ket{3}}{\sqrt{2}}$, $\ket{3}$ respectively. The real part of the displacement is shown, plus the measurement controls in the superposition case. As expected, when the input corresponds to the target state, all neural networks output zero drive, since the correct state has been reached. Interestingly, when the target state is a pure Fock state, the neural network decides to displace only slightly (i.e. $\beta(t)\sim 0$ ) when the input state is a superposition between the target state and another one. In Fig.~\ref{fig:figure5}a we show the measurement strategy when the target is a superposition, $\frac{\ket{1}+\ket{3}}{\sqrt{2}}$. The network measures consistently state $\ket{2}$, independently from the input state, matching what was already shown in Fig.~\ref{fig:figure4}.

It is equally interesting to analyze the dependence of the discovered strategy on the physical parameters, i.e. the measurement rate $\Gamma_{\rm meas}$ and the maximum available drive strength $\beta_{\rm max}$. The results are displayed in Fig.~\ref{fig:figure5}c.
While we sweep through different parameter values, each network is trained completely anew by our RL approach, but in principle one could start from an already trained network on a different set of parameters and use that as initialization for the new training. This strategy would save computation time but its adoption would not change the results shown in Fig.~\ref{fig:figure5}. We observe that the RL approach yields high fidelities once both the maximum allowed drive and the measurement rate exceed a certain threshold. Specifically, both the drive $\beta_{\rm max}$ and $\gamma_{\rm meas}$  need to be multiplied by the total time span $T_{\rm max}$ allowed for the control task, and the resulting dimensionless numbers need to cross a certain threshold for each of these quantities, as seen in the figure.
In addition, we have compared the RL results with the greedy algorithm (sketched in Appendix A). It can be readily seen in Fig.~\ref{fig:figure5}b that the RL algorithm can reach higher average fidelities in a wider ranges of parameters, and performs slightly worse only when the maximum displacement allowed is quite small.

Finally, we turn our attention to check the robustness of the reinforcement learning strategies to different kinds of noises. Specifically, we implemented two different disturbances of the system: decay and dephasing. The decay is modelled with an additional term in the master equation \eqref{eq:SME_final} of the form $\gamma_{\rm decay} {\mathcal D}[{\hat a}]$.  Dephasing is covered by a term $\dfrac{\gamma_{\rm dephasing}}{2} \sum_n {\mathcal D}[{\hat P}_n]$. We note that this additional dephasing can also be seen as describing an inefficient measurement, without any corresponding increase of information (this results in an impure quantum state). The results of this analysis are shown in Fig.~\ref{fig:figure6}, where the RL approach has been used to discover strategies for different amounts of decay and dephasing. The RL strategy can perform well even when both kinds of noises are present in the system, and it outperforms the greedy strategy. While the deleterious influence of dephasing is comparatively small, the effects of decay seem to be significantly more severe (when comparing results at the same rate). This insight provides an important guideline for future experiments.

\section{Conclusion}

The present work has demonstrated the feasibility of using deep RL to compensate limited control by nonlinear measurements in an experimentally relevant setting. However, it also has underlined the importance of fundamental questions about controllability under feedback: in our scenario, under some circumstances, RL can only reach a limited fidelity, but is this a necessary consequence of the limited control or could it be overcome by better strategies that the adopted RL approach simply did not discover?

The experimental implementation of the strategies discovered here, for quantum state preparation in a cavity, will rely either on fast on-chip processing to extract the quantum state update from the observed measurement signal or will exploit the general idea of two-stage learning, training a deployable network in a supervised fashion after RL discovery of the strategy has succeeded. It would be equally interesting to see whether direct RL training starting from the measurement trajectories obtained in the experiment could work, although that will likely require improved algorithms.

In the future, the approach adopted here could be applied to other scenarios such as cavities with nonlinearities where the control consists either in a drive (as in the present work), in a detuning, or even in a time-dependent control of the strength of the nonlinearity. These control strategies are well suited to quantum error correction based on bosonic codes~\cite{caiBosonicQuantumError2021}, and could thus be instrumental in the development of a fault tolerant quantum processor. The extension to quantum many-body scenarios, like multiple coupled nonlinear cavities or qubits, seems equally promising. In all these cases, it would be hard to find suitable quantum feedback strategies without the help of a tool as powerful as deep RL.

\section{Data availability}
Data and code are available from the corresponding author on reasonable request.

\bibliographystyle{quantum}
\bibliography{bibliography.bib}

\onecolumn\newpage
\appendix

\section*{Appendix A}
\label{section:appendixA}
\begin{figure}[t]
\centering
	\includegraphics[width=0.5\linewidth]{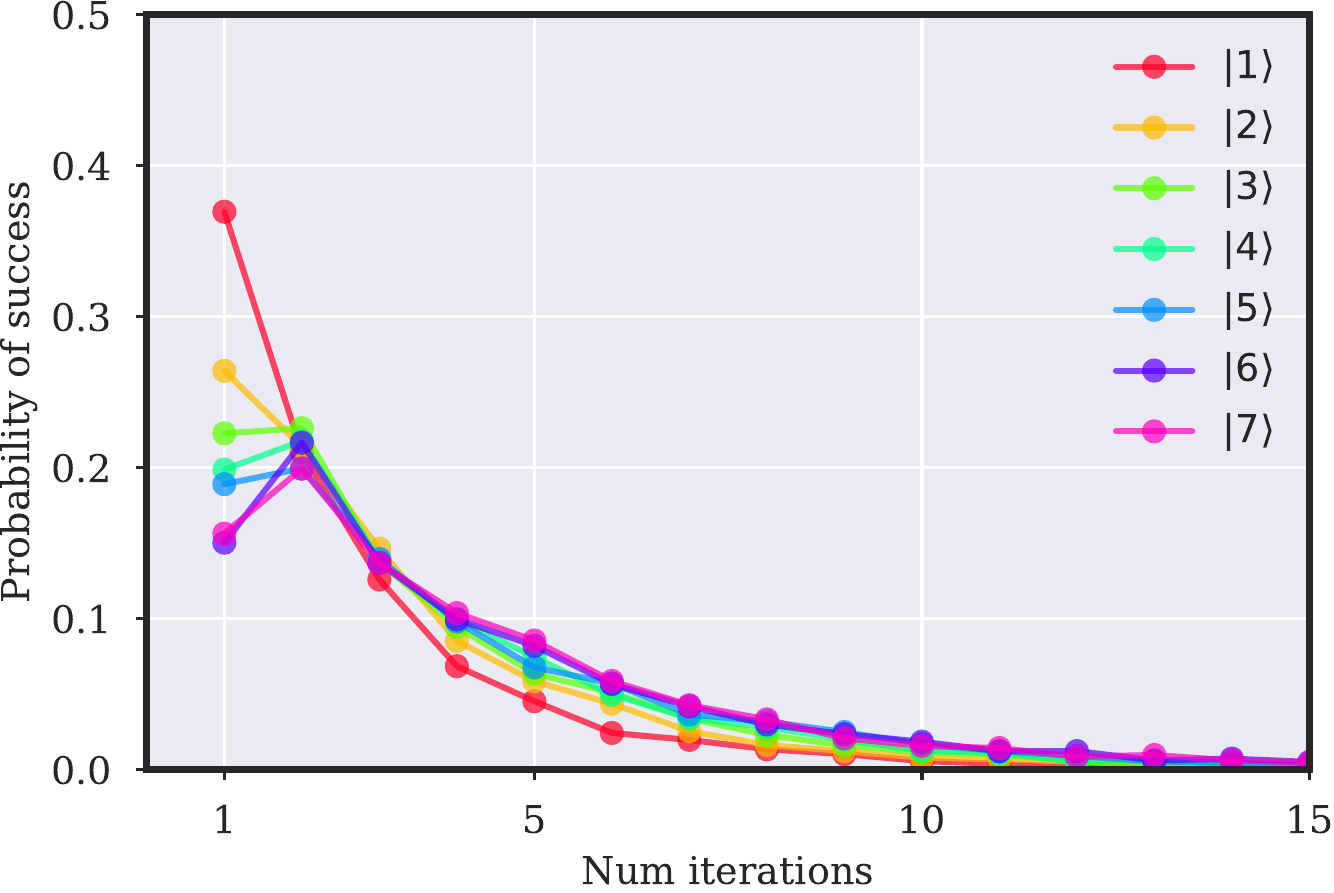} 
	\caption{
Strong measurement-based strategy applied for first 7 Fock states. The probability of success is not monotonic for the first two iterations, then decreases. The cut-off Fock state is $\ket{70}$, since we empirically proven that the probability of displacing higher states to obtain a Fock state with $n<7$ will basically converge to $0$. It is worth noticing that the cumulative sum of these probabilities, for each target Fock states, never goes above 0.91, denoting a probability of not succeeding with this strategy in a finite time which is close to $9\%$.
    }
	\label{fig:comparison}
\end{figure}

In the Fock state basis, the displacement operator acts on $|n\rangle$ like \cite{deoliveiraPropertiesDisplacedNumber1990,nietoDisplacedSqueezedNumber1997}:

\begin{equation}
D(\alpha)|n\rangle = \exp[-|\alpha|^2/2]\sum_{k=0}^{\infty} \frac{\alpha^k}{k!}
   \sum_{j=0}^{n}\frac{(-\alpha^*)^j}{j!}
   \left[\frac{(n-j+k)!n!}{(n-j)!(n-j)!}\right]^{1/2}|n-j+k\rangle.
\end{equation}
Ideally, to drive a system from $|n\rangle$ to $|l\rangle$, one wants to maximize:

\begin{equation}
\langle l| D^{\dagger}_{\alpha} \rho  D_{\alpha} |l \rangle
\end{equation}
with $\rho=|n\rangle \langle n|$.
By introducing two identities:

\begin{equation}
\label{"start"}
\langle l| D^{\dagger}_{\alpha} \rho  D_{\alpha} |l \rangle = \sum_{l',m'} \langle l| D^{\dagger}_{\alpha} |l'\rangle\langle l'| \rho |m'\rangle\langle m'| D_{\alpha} |l \rangle
\end{equation}

So we need to compute $\langle l|  D_{\alpha} |n \rangle$ for a generic $n$ and $l$.
\begin{align}
    \langle l|  D_{\alpha} |n \rangle &=\exp[-|\alpha|^2/2]\sum_{j=0}^{n} \sum_{k=0}^{\infty} \frac{\alpha^k}{k!}
   \frac{(-\alpha^*)^j}{j!}
   \left[\frac{(n-j+k)!n!}{(n-j)!(n-j)!}\right]^{1/2} \overbrace{\langle l|n-j+k\rangle}^{\delta_{l,n-j+k}}\\
   &=\exp[-|\alpha|^2/2]\sum_{j=max(0,n-l)}^{n}  \frac{\alpha^{-n+j+l}}{(-n+j+l)!}
   \frac{(-\alpha^*)^j}{j!(n-j)!}
   \sqrt{l! n!} \\
   &=\exp[-|\alpha|^2/2] \sqrt{l! n!}  \alpha^{-n+l} \sum_{j=max(0,n-l)}^{n}  \frac{|\alpha|^{2j}}{  (-n+j+l)!j!(n-j)! }
\end{align}
Plugging this into Eq.~\eqref{"start"},and assuming $\rho=|n\rangle \langle n|$:
\begin{equation}
\sum_{l',m'} \langle l| D^{\dagger}_{\alpha} |l'\rangle\overbrace{\langle l'| n\rangle}^{\delta_{l',n}}\overbrace{ \langle n |m'\rangle}^{\delta_{n,m'}}\langle m'| D_{\alpha} |l \rangle= \langle l| D^{\dagger}_{\alpha} |n\rangle\langle n| D_{\alpha} |l \rangle=\left| \langle l| D^{\dagger}_{\alpha} |n\rangle \right|^2
\end{equation}
Finally:
\begin{equation}
\left| \langle l| D^{\dagger}_{\alpha} |n\rangle \right|^2=\left| \exp[-|\alpha|^2]l!n!\alpha^{2(l-n)} \sum_{j=max(0,n-l)}^{n}  \frac{|\alpha|^{2j}}{  (-n+j+l)!j!(n-j)! } \right|^2
\end{equation}
As example, we can compute this overlap with $\ket{\psi_i}=\ket{i}=\ket{0}$ and  $\ket{\psi_f}=\ket{l}=\ket{2}$, obtaining:
\begin{equation}
\langle 2| D_{\alpha} |0\rangle \langle 0|  D^{\dagger}_{\alpha} |2 \rangle=\frac{1}{2} e^{-\alpha^2} \alpha^4 
\end{equation}We now can compute the optimal $\alpha$ that maximizes the overlap between any initial state $\ket{n}$ and target $\ket{l}$, namely $\alpha_{optim}(n, l)$. After computing this table of values, we can develop a strategy to achieve a particular target Fock state $\ket{l}$. The idea is is to start from $n=0$ (i.e. the ground state) and then we displace by the optimal  $\alpha_{optim}(0, l)$. Then, we strong measure the evolved state, and this will collapse to a particular Fock state $\ket{l_1}$. If $l_1=l$, then the algorithm ends and the trajectory was successful. If, instead, $l_1\neq l$ , we displace $\ket{l_1}$ by the optimal $\alpha_{optim}(l_1, l)$ and we iterate this project until we reach $\ket{l}$ or the algorithm reaches a very high Fock state (discussed in Fig.~\ref{fig:comparison}). We use this strategy for each target Fock state from $n=1$ to $n=7$, as shown in Fig.~\ref{fig:comparison}. We see that the probability of reaching the target state is very high in the first few iterations, while it slowly decreases otherwise.

Now we will try to find an optimal strategy that doesn't assume Fock states as target states This is referred in the main text as "greedy strategy". \\
The system starts in the initial state $\ket{\psi_i}$. Then, this state is evolved through the master equation describing the system in hand, but with every possible value of the displacement $D(\alpha)$. We discretise the values of $\alpha$ in $N$ possible values, therefore obtaining $N$ different target states after evolving $\ket{\psi}(t)$ to $\ket{\psi}(t+\Delta t)$. Among these $N$ states we take the one that maximises the fidelity with the target state that we want, i.e. $\ket{\psi_f}$. We then repeat this procedure for every timestep in a trajectory. This strategy is referred to as "greedy", since the best possible strategy could not pass through local optima, but in some cases an optimal strategy would rely in a completely different path. \\

\section*{Appendix B: PPO}
\label{section:appendixB}
PPO \cite{schulmanProximalPolicyOptimization2017} is an on-policy algorithm, meaning that the policy is updated directly from the data and it doesn't need to collect them in order to update it, like off-policy algorithms do. This means that PPO could be used directly attached to an experimental setup, as opposed to off-policy algorithms (like Q-learning). The main idea behind PPO is a policy gradient method, which consists in computing an estimator of the policy gradient and then plugging it in a stochastic gradient ascent algorithm. Such estimator is defined as:
\begin{equation}
\label{loss}
L(\theta) = \hat{\mathrm{E}}_t \left[  \log \pi_{\theta}(a_t|s_t) \hat{A}_t\right]
\end{equation}
where $\pi_{\theta}(a_t|s_t)$ is the probability of choosing action $a_t$, given state $s_t$ and according the policy parameters $\theta$. $\hat{A}_t$ is called advantage function and estimates the value (i.e. the cumulative reward from timestep $t$ onwards) of action $a_t$. $\hat{A}_t$ is then composed of two terms:
\begin{equation}
    \hat{A}_t = \sum_{t'=t}^{t_{max}} \gamma^{t'-t} r_t - \hat{V}(s_t)
\end{equation}
where the first one represents the (discounted) some of rewards (i.e. discounted return) from timestep $t$ and the second one a value function that estimates the discounted reward. If $\hat{A}_t>0$ this means that action $a_t$ would collect more reward as expected and the agents needs to select it more often in future episodes. To estimate $\hat{V}(s_t)$ we use a second neural network, which takes state $s_t$ as input, but one could use in principle only the policy neural network and branch the last layer and adding one additional continuous-valued neuron, exploiting the already processed state after the first layers of the network. The problem with estimator Eq.~\eqref{loss} is that multiple steps of optimization using the same trajectory could bring to potentially too big policy updates. The idea of PPO is to optimize an objective function that constrains its update. This objective function is defined as:
\begin{equation}
L^{PPO}(\theta) = \hat{\mathrm{E}}_t \left[   \min(r_t(\theta)) \hat{A}_t, \text{clip}(r_t(\theta), 1-\varepsilon, 1+\varepsilon)\hat{A}_t)        \right]
\end{equation}where $r_t(\theta)$ is defined as:
\begin{equation}
r_t(\theta) = \frac{\pi_{\theta}(a_t|s_t) }{\pi_{\theta_{\rm old}}(a_t|s_t) },
\end{equation}$\theta_{\rm old}$ are the weights of the policy before the update and $\varepsilon$ is an hyperparameter.

We implemented the PPO algorithm by using the library Stable Baselines \mbox{\cite{hillStableBaselines2018}}. The neural network used is a multi-layer perceptron with two hidden layers of size 64 and tanh activation function. The hyperparameters used for the PPO implementation are the following:
\begin{table}[h]
\centering
\label{tab:PPO}
\begin{tabular}{|c|c|}

\hline
Parameter&Value\\
\hline
gamma & 0.99\\
n\_steps & 128\\
clip\_param & 0.2\\
ent\_coef & 0.0\\
learning\_rate & 0.00025\\
vf\_coef & 0.5\\
max\_grad\_norm & 0.5\\
lam & 0.95\\
nminibatches & 4\\
noptepochs & 4\\
cliprange & 0.2\\
adam\_epsilon & 0.00001\\
optim\_stepsize & 0.001\\
optim\_batchsize & 64\\
timesteps\_per\_actorbatch & 256\\
\hline
\end{tabular}
\end{table}

\end{document}